\documentclass[twocolumn,amsmath,amssymb,pra]{revtex4}
\usepackage{graphicx}
\usepackage{dcolumn}
\usepackage{bm}
\usepackage{subfigure}
\newcounter{one}
\newcommand{\op}[1]{\hat {#1}}

\newcommand{\hatd}[1]{{\hat {#1}^\dagger}}
\def\vect#1{\mbox{\boldmath $#1$}}

\newcommand{\ket}[1]{| #1 \rangle}

\DeclareMathOperator{\arctanh}{arctanh}

\begin{document}

\title{Universal linear Bogoliubov transformations through one-way quantum computation}

\author
{Ryuji Ukai$^{1}$, Jun-ichi Yoshikawa$^{1}$, Noriaki Iwata$^{1}$, Peter van Loock$^{2}$, and Akira Furusawa$^{1}$}

\affiliation{$^{1}$Department of Applied Physics and Quantum-Phase Electronics Center, School of Engineering, The University of Tokyo, 7-3-1 Hongo, Bunkyo-ku, Tokyo 113-8656, Japan\\
$^{2}$  Optical Quantum Information Theory Group,
Max Planck Institute for the Science of Light, \\
Institute of Theoretical Physics I, Universit\"{a}t Erlangen-N\"{u}rnberg,
Staudtstr.7/B2, 91058 Erlangen, Germany}

\begin{abstract}
We show explicitly how to realize an arbitrary linear unitary Bogoliubov
transformation (LUBO) on a multi-mode quantum state through homodyne-based one-way quantum computation.
Any LUBO can be approximated by means of a fixed, finite-sized, sufficiently squeezed
Gaussian cluster state that allows for
the implementation of beam splitters (in form of three-mode connection gates)
and general one-mode LUBOs. In particular, we demonstrate that
a linear four-mode cluster state is a sufficient resource for an arbitrary one-mode LUBO.
Arbitrary input quantum states including non-Gaussian states
could be efficiently attached to the cluster through quantum teleportation.
\end{abstract}

\maketitle

\section{INTRODUCTION}

The cluster model of quantum computation, or one-way quantum computation \cite{Raussendorf01,Briegel01},
is an alternative approach to the standard circuit model for quantum computing \cite{NielsenChuang}.
In the cluster model, a special type of entangled state is used as a resource for cluster computation.
These resource states are known as cluster states.
A cluster computation is basically a sequence of elementary, `half' teleportations \cite{Nielsen06,Zhou00} where quantum information is not only transmitted through a cluster state but also manipulated in any desired way depending on the specific choice of the measurement bases
at each teleportation step. As opposed to standard-teleportation-based schemes,
the measurements in a cluster computation are all local (subsequently performed on the individual nodes of the cluster).
In order to achieve universal quantum computation using a fixed cluster state, active feedforward is needed, where the measurement bases of subsequent measurements have to be adjusted according to the outcomes of the earlier measurements.

Cluster states and cluster computation were originally proposed for discrete variables (DV), namely qubits \cite{Raussendorf01,Briegel01}.
More recently, the cluster-state model was then extended to the regime of continuous variables (CV) \cite{Zhang06, Menicucci06},
in which universal cluster states can be approximated by experimentally highly accessible Gaussian multi-mode squeezed states of sufficiently many
quantized optical modes (qumodes).
Both for DV and for CV, the cluster-state model is known to be equivalent to the circuit model in the sense that
any finite-dimensional (qubits) as well as any infinite-dimensional (qumodes) operation can be efficiently realized
in a cluster-based scheme.

For DV, an arbitrary single-qubit rotation (unitary) can be exactly decomposed into three elementary single-qubit rotations \cite{NielsenChuang}.
Therefore, even though the whole set of single-qubit unitaries is continuous, concatenating three elementary (but continuous) single-qubit rotations in a three-step cluster computation using a linear four-qubit cluster state is sufficient to achieve universality in the single-qubit space.
Such elementary rotations by general angles would include so-called non-Clifford gates; in this case, feedforward is required during
the cluster computation. As a result, provided that the continuous, elementary single-qubit rotations can be implemented
in an error-resistant fashion, any multi-qubit unitary can be performed by connecting sufficiently many linear four-qubit clusters
by vertical wires through which a fixed two-qubit entangling gate can be applied when needed.

In the case of CV, there are various subtleties, even in theory. First, independent of the cluster model,
an arbitrary single-qumode transformation (represented by a Hamiltonian which is an arbitrary polynomial of
the qumode's position and momentum variables) must include (arbitrary) higher-order, nonlinear (non-Gaussian) transformations \cite{detail2}.
For this purpose, full universality has been shown to be asymptotically approachable through infinite
(but efficient) concatenation of
a finite set of elementary unitaries, each lying in the neighborhood of the identity, and including at least
one nonlinear gate \cite{Lloyd99}.

Secondly, when utilizing cluster states,
in order to satisfy the above notion of full universality for CV, sufficiently large (potentially infinite) squeezing of the Gaussian
cluster state is required, as otherwise the asymptotic concatenation of elementary gate teleportations
would accumulate an infinite amount of finite-squeezing-induced errors.
The second issue here, the issue of finite squeezing, is then related with the first issue, the issue of full universality
for CV based on infinite, elementary-gate concatenation. Although it has been proven that the squeezing per mode needed to create
a universal Gaussian cluster state of fixed accuracy does not depend on the size of the cluster state (and hence on the size of the
computation it is used for) \cite{Gu09}, the errors in a cluster computation using a fixed-accuracy cluster would nonetheless grow
arbitrarily with the length of the computation (and the size of the cluster).

In this paper, we focus on a restricted class of cluster computations, namely those realizing linear, Gaussian transformations
corresponding to quadratic Hamiltonians. More generally,
these transformations are referred to as linear unitary Bogoliubov (LUBO) transformations.
In this case, it is well-known that arbitrary quadratic Hamiltonians can be exactly and finitely decomposed into elementary
quantum optical elements such as single-mode squeezers and beam splitters \cite{Reck94,Braunstein05}.
A perfect simulation of the total Hamiltonian no longer requires an infinite concatenation of these elementary optical gates;
each elementary gate no longer has to be weak and may even be far from the identity.
These properties greatly simplify the theoretical analysis and the experimental implementation
of LUBO transformations through cluster computation over CV. As the Gaussian transformations play the roles of the Clifford gates
for CV, the measurements in a Gaussian cluster computation may all be done in parallel (`Gaussian parallelism');
moreover, local homodyne detections on the individual qumodes of the cluster are sufficient to achieve any multi-mode LUBO transformation \cite{Menicucci06}.

Despite these known simplifications and possibly because of the known impossibility of full universality
in the case of Gaussian cluster computations, so far there has been no explicit derivation
of universal cluster states for Gaussian/Clifford computations which would include an explicit choice of homodyne measurements
on a specifically shaped finite-sized cluster state realizing operations far from the identity.
It has only been shown how a single-mode squeezing transformation can be approximately applied to an arbitrary input
state attached to a perfect (infinitely squeezed), linear four-mode cluster state \cite{Peter07J}.

Here we shall give several such explicit derivations. In particular,
we show that an arbitrary one-mode LUBO transformation can be {\it perfectly} achieved through an ideal four-mode linear cluster state.
Further, we show that an arbitrary input state can be coupled to the cluster state using standard quantum teleportation
\cite{Vaidman94,Braunstein98}.
Finally, we present a simple idea that enables one to implement an arbitrary multi-mode Gaussian transformation.
Even though we will not give a provably optimal, multi-mode solution with regard to the size of the cluster,
in our proposed scheme, the dependence of the cluster size is quadratic on the number of the input modes and
this order coincides with the minimum order of elements required for general multi-mode Gaussian transformations.

As a consequence of our results, the efficient experimental implementation of {\it any} multi-mode
LUBO transformation on {\it any} optical multi-mode quantum state (especially including non-Gaussian input states) becomes
possible using the existing optical schemes for efficient, deterministic creation of Gaussian cluster states
\cite{Peter07P,Su07,Yukawa08,Meni07,Meni08}.
In other words, the entire regime of multi-mode linear optical transformations becomes, in principle, accessible through
one fixed, offline squeezed, finite-sized cluster state and homodyne detections on it.

The plan of the paper is as follows. First, in
Sec.~\ref{elemgatetelep}, we will give a brief introduction into
cluster computation over CV including the elementary teleportation
circuits for gate teleportation. In Sec.~\ref{onemodeLUBO},
we explicitly derive the linear four-mode cluster state
and the homodyne measurement steps which allow for a realization
of arbitrary one-mode LUBO transformations. In order to attach
arbitrary quantum states to the cluster in an efficient way,
we show in Sec.~\ref{telepcoupling} how one may employ standard
quantum teleportation for this purpose.
An explicit scheme for a one-mode LUBO transformation
using teleportation-based input-cluster coupling is discussed
in Sec.~\ref{onemodeLUBOwtelep}.
Finally, before concluding in Sec.~\ref{conclusion},
we examine the most general case of universal multi-mode
LUBO transformations in Sec.~\ref{multimodeLUBO}.

\section{ELEMENTARY GATE TELEPORTATIONS}\label{elemgatetelep}
Before going into detail, we shall briefly review the basic concepts of continuous-variable (CV) cluster computation in quantum optics.
We use the convention $\hbar=1/2$ such that $[\hat{x},\hat{p}] = i/2$ for $\hat{a}=\hat{x}+i\hat{p}$
and $[\hat{a},\hat{a}^\dagger] = 1$, where
the real and imaginary parts of an optical qumode's annihilation operator are as usual expressed by the
position and momentum operators $\hat{x}$ and $\hat{p}$, respectively.

\begin{figure}[b]
\centering
\includegraphics[width=4cm,clip]{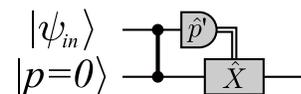}
\caption{An elementary one-mode one-way QC gate;
$\ket{\psi_{in}}$ is the input state; $\ket{p=0}$ is a momentum eigenstate with eigenvalue zero;
$\op{p}^\prime$ is the measurement variable and $\op{X}$ a correction displacement operator.
}
\label{FigOneStepTele}
\end{figure}

The building block of a one-mode cluster computation is shown %with a dashed box
in Fig.~\ref{FigOneStepTele}. It can be considered as a generalized (kind of `half') teleportation
\cite{Nielsen06,Zhou00}.
First, the input state $\ket{\psi}$ and an ancilla squeezed vacuum state $\ket{p=0}$ are coupled through a
CV quantum nondemolition (QND) interaction.
A QND coupling between modes $j$ and $k$ is described by the gate $\exp(2i\hat{x}_j\hat{x}_k)$, which is depicted in Fig.~\ref{FigOneStepTele} as a line that connects the two horizontal wires for each qumode.
Next, the input mode is subject to a local measurement with a measurement basis $\{\hat{O}^\dagger \ket{p}\}$ (that is, the measured observable is $\hat{p}^\prime=\hat{O}^\dagger\hat{p}\hat{O}$), where $\hat{O}$ is a function of only $\hat{x}$, i.e., $\hat{O}=\exp[i f(\hat{x})]$.
After the feedforward operation $\hat{X}_j(s)=\exp(-2i s\hat{p}_j)$, which is a position displacement in phase space by the value of the measurement outcome $s$, the resulting output state corresponds to $\ket{\psi^\prime}=\hat{F}\hat{O}\ket{\psi}$, where $\hat{F}=\exp[-i(\pi/2)\hatd{a}\hat{a}]$ is the Fourier transform
operator.
In the realistic case, $\ket{p=0}$ will be approximated by a single-mode finitely squeezed state.
As a result, some unwanted excess noise is introduced at each teleportation step of the computation
depending on the initial squeezing level.

\begin{figure}
\centering
\includegraphics[width=7.5cm,clip]{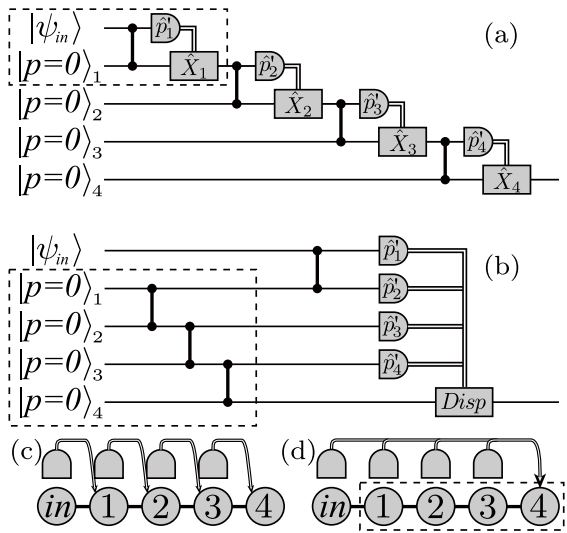}
\caption{(a)~One-step elementary one-way QC gate (enclosed by a dashed line)
and its cascade (the whole).
(b)~An equivalent circuit to (a). The circuit enclosed by a dashed line shows a four-mode linear cluster state.
(c) and (d) are graph representations of (a) and (b), respectively.
Each ball shows a single mode, while each thick line shows a QND connection.
A dashed box in (d) is a four-mode linear cluster state which corresponds to that in (b).
}
\label{FigFourStepTele}
\end{figure}

Arbitrary one-mode transformations can then be performed by concatenating sufficiently many
elementary teleportation steps.
Similarly, when several modes propagate through a two-dimensional cluster state
(such as a 2D lattice), QND gates can be applied to any two modes during the cluster
computation such that universal multi-mode transformations become possible
\cite{Menicucci06}.

Fig.~\ref{FigFourStepTele} (a) shows an example of a cascade of teleportation steps
for one-mode manipulations.
Every single step $i$ will apply the operation $\op{F}\op{O}_i$. Hence
the general output state of an $n$-time cascaded one-mode circuit corresponds to
\begin{align}
&\hat{F}\hat{O}_n(\hat{x})\dots\hat{F}\hat{O}_3(\hat{x})\hat{F}\hat{O}_2(\hat{x})\hat{F}\hat{O}_1(\hat{x})\ket{\psi}_\text{in} \notag\\
&=\hat{F}\dots\hat{F}\hat{O}_n(\cdot)\dots\hat{O}_3(-\hat{x})\hat{O}_2(\hat{p})\op{O}_1(\hat{x})\ket{\psi}_\text{in}.
\end{align}
As one can see, elementary unitary operations, either diagonal in $\hat{x}$ or in $\hat{p}$, are alternately performed on the input state.

One important thing here is that the QND coupling $\exp(2i\hat{x}_j\hat{x}_k)$ is an element of the Clifford group $C_2$, which is a group
that consists of the normalizers of the Heisenberg-Weyl (HW) group $C_1$,  i.e., \ $C_2=\{\hat{U}|\hat{U}C_1\hat{U}^\dagger = C_1\}$.
The HW group $C_1$ is the group of phase space displacements, an element of which is generally written in the form $\exp[2i\sum_j(\eta_j\hat{x}_j-\xi_j\hat{p}_j)+i\phi]$ where
$\eta_j$ and $\xi_j$ are arbitrary real values that represent the size of the displacements in phase space for mode $j$, and
$\phi$ is a global phase.
The Clifford group $C_2$ is a group whose generators are polynomials up to quadratic order in position $\hat{x}_j$ and momentum $\hat{p}_j$, i.e., the elements take on the general form $\exp[i\sum_{j,k}(\alpha_{j,k}\hat{x}_j\hat{x}_k+\beta_{j,k}\hat{x}_j\hat{p}_k+\gamma_{j,k}\hat{p}_j\hat{p}_k)+i\sum_{j}(\delta_j\hat{x}_j+\epsilon_j\hat{p}_j)+i\phi]$,
where $\alpha_{j,k}$, $\beta_{j,k}$, $\gamma_{j,k}$, $\delta_j$, and $\epsilon_j$ are arbitrary real values.

As a consequence of the discussion of the preceding paragraph,
all the QND couplings can be applied prior to an actual quantum computation, while the feedforward operations remain simple
displacements in phase space [Fig.~\ref{FigFourStepTele}(b)].
The resulting multi-mode entangled state [see the dashed box in Fig.~\ref{FigFourStepTele}(b)], in which several single-mode squeezed states are coupled  through pairwise QND interactions, is the resource cluster state.

In the following, a cluster state built from `blank' squeezed vacuum modes (i.e., without an input quantum state attached to it)
shall be referred to as an ``ancilla cluster state''.
Once such a resource state has been prepared, the individual displacements of every teleportation step can then all be postponed
until the end of the cluster computation, as illustrated in Fig.~\ref{FigFourStepTele}(b).
However, it does make a difference whether the desired operation $\op{O}_j \in C_2$ or $\op{O}_j \notin C_2$.
In the latter case, when $\op{O}_j \notin C_2$ for some $j$ (corresponding to cubic or higher-order gates),
the measurement bases of the succeeding ($(j+1)$th, $(j+2)$th, $\dots$)
teleportation steps would depend on the outcome of measurement $j$.
More conveniently, when $\op{O}_j \in C_2$ for all $j$, none of the chosen measurement bases depend on any measurement outcomes
such that all the measurements can be performed in any order.

Cluster states are often represented using graphs \cite{Hein04}, as, for example, the four-mode linear chain in the dashed box of Fig.~\ref{FigFourStepTele}(d) where each node denotes an ancilla single-mode squeezed state and each link represents a QND coupling.
Using such graphs, we can easily distinguish different types of entangled cluster states.
A perfect cluster state can be approached in the limit of infinite ancilla squeezing with the resulting
quantum correlations for all $j$ \cite{Peter07P},
\begin{align}
\hat{p}_j-\sum_{k\in N(j)}\hat{x}_k\rightarrow 0,
\end{align}
where $N(j)$ denotes the set of all nearest neighbors to the $j$-th mode. In the limit of infinite squeezing,
these quantum correlations among the qumodes' quadratures uniquely determine the corresponding graph state.
The correlations are analogous to the generators of the stabilizer group for a qubit graph state \cite{Peter07P}.
The only difference here is that for CV, it is more convenient to express the stabilizer conditions
in terms of the Lie algebra, i.e., the generators of the HW Lie group, for which the stabilizers become
`nullifiers' \cite{Gu09}.

In the following, we restrict ourselves to unitary Gaussian transformations on $n$ modes,
which form a Clifford group $C_2=\text{Cl}(n)$. The Clifford group is a semidirect product of the symplectic group $\text{Sp}(2n, \mathbb{R})$ and the HW group $C_1=\text{HW}(n)$, $\text{Cl}(n) = \text{Sp}(2n, \mathbb{R})\ltimes \text{HW}(n)$.
The group $\text{HW}(n)$ is a homogeneous space under the adjoint action of $\text{Cl}(n)$, and one can construct a group representation of $\text{Cl}(n)$ on the vector space of its Lie algebra $\text{hw}(n)$.
Here, instead of using this particular representation, we prefer to consider a representation isomorphic to the former one,
but revealing a clearer physical meaning: the linear transformation of position $\hat{x}$ and momentum $\hat{p}$ in the
Heisenberg picture,
\begin{equation}\label{LUBOinxandp}
\begin{pmatrix}
\vect{\hat{x}}^\prime \\
\vect{\hat{p}}^\prime
\end{pmatrix}
=
\hat{U}_{G(n)}^\dagger
\begin{pmatrix}
\vect{\hat{x}} \\
\vect{\hat{p}}
\end{pmatrix}
\hat{U}_{G(n)}
=
\begin{pmatrix}
A & B \\
C & D
\end{pmatrix}
\begin{pmatrix}
\vect{\hat{x}} \\
\vect{\hat{p}}
\end{pmatrix}
+
\begin{pmatrix}
\vect{e} \\
\vect{f}
\end{pmatrix},
\end{equation}
where $\vect{\hat{x}}$ ($\vect{\hat{x}}^\prime$) and $\vect{\hat{p}}$ ($\vect{\hat{p}}^\prime$) denote the vectors of position and momentum operators $\vect{\hat{x}} = (\hat{x}_1, \dots, \hat{x}_n)^T$ and
$ \vect{\hat{p}} = (\hat{p}_1, \dots, \hat{p}_n)^T$ at the input (output), respectively.
The $2n \times 2n$ matrix $M_{G(n)} = \left(\begin{smallmatrix} A & B \\C & D\end{smallmatrix}\right)$ is a faithful representation of the symplectic group $\text{Sp}(2n, \mathbb{R})$ with $2n^2+n$ degrees of freedom.
Here, the matrix $M_{G(n)}$ is divided into four $n\times n$ matrices $A$, $B$, $C$, and $D$.
The column vectors $\vect{e}, \vect{f}\in \mathbb{R}^{n}$ represent displacements in phase space.
The isotropy subgroup of this representation is a global phase $\exp(i\phi)$ which we can ignore.
The displacements will be omitted as well, as they can be trivially applied at any time
during a cluster computation \cite{Menicucci06,Peter07J}.
Note that Eq.~(\ref{LUBOinxandp}) corresponds to an $n$-mode LUBO transformation,
usually expressed in terms of annihilation and creation operators,
$\hat a_k = \sum_{l} \tilde A_{kl} \hat a_l +  \tilde B_{kl} \hat a_l^\dagger + \gamma_k$,
with the $\gamma_k$ being $n$ complex parameters and the $n \times n$ matrices
$\tilde A$ and $\tilde B$ chosen such that the bosonic commutators
are preserved.

\section{UNIVERSAL ONE-MODE LUBO}\label{onemodeLUBO}
Let us now start with the explicit realization of an arbitrary one-mode Gaussian transformation $M_{G(1)} = \left(\begin{smallmatrix} a & b \\c & d \end{smallmatrix}\right)$, where $a d-b c=1$.
In cluster computation, the elementary gate for one-mode LUBO/Gaussian transformations is the quadratic phase gate $\hat{O}_G(\hat{x})=\exp(i\kappa\hat{x}^2)$ \cite{Miwa09}, where $\kappa$ takes on arbitrary real values,
together with the Fourier transform.
Therefore, our strategy will be to search for decompositions of a given LUBO transformation into quadratic $x$ and $p$ phase gates.
In case of the $x$ phase gate, the corresponding observable to be measured is $e^{-i\kappa\hat{x}^2}\hat{p}e^{i\kappa\hat{x}^2} =\hat{p}+\kappa\hat{x} =g(\hat{x}\sin\theta+\hat{p}\cos\theta)$, where $g=\sqrt{1+\kappa^2}$ and $\theta=\arctan\kappa$.
In an optical implementation, any such linear combination of $\hat x$ and $\hat p$
can be measured by means of homodyne detection with a suitable choice of the local oscillator phase
depending on the angle $\theta$.

The 2$\times$2 matrix representation of $\hat{O}_G(\hat{x})$ is $O(\kappa)=\left(\begin{smallmatrix} 1 & 0 \\ \kappa & 1 \end{smallmatrix}\right)$ and that of the Fourier transform $\hat{F}$ is $F=\left(\begin{smallmatrix} 0 & -1 \\ 1 & 0 \end{smallmatrix}\right)=R(\pi/2)$ where $R(\theta)=\left(\begin{smallmatrix} \cos\theta & -\sin\theta \\ \sin\theta & \cos\theta \end{smallmatrix}\right)$ is a phase space rotation.
Thus, the total transformation of a one-step one-mode teleportation gate becomes $M(\kappa)=FO(\kappa)=\left(\begin{smallmatrix} -\kappa & -1 \\ 1 & 0 \end{smallmatrix}\right)$.

Now first we prove the following lemma.
This lemma will also be useful later on, so we describe it in a somewhat general form.\\
{\it Lemma}:~Let us combine those one-mode Clifford group ($C_2$)
operations that are performed before the final elementary gate
$M(\kappa_n)$ into $\mathbf{M}_{n-1}(\kappa_1,\dots,\kappa_{n-1}) = \left(\begin{smallmatrix} a_{n-1} & b_{n-1} \\ c_{n-1} & d_{n-1} \end{smallmatrix}\right)\in\text{Sp}(2, \mathbb{R})$ where $(\kappa_1,\dots,\kappa_{n-1})$ are the free parameters in the choice of the measurement bases.
Then, {\it together with} the final step $M(\kappa_n)=FO(\kappa_n)$, an arbitrary one-mode $C_2$ operation
is accomplished if and only if (iff) $(a_{n-1}, b_{n-1})\in\mathbb{R}^2$ covers the whole range of $\mathbb{R}^2\setminus\{0,0\}$.
This means that a certain property of the whole circuit without the last step, $\mathbf{M}_{n-1}$, determines whether
the circuit as a whole is universal or not.\\
{\it Proof}: The matrix representation after the final step can be written as,
\begin{align}
\mathbf{M}_{n}(\kappa_1,\dots,\kappa_{n})&\equiv M(\kappa_n)\mathbf{M}_{n-1}(\kappa_1,\dots,\kappa_{n-1}) \notag\\
&=
\begin{pmatrix}
-c_{n-1}-\kappa_n a_{n-1} & -d_{n-1}-\kappa_n b_{n-1} \\
a_{n-1} & b_{n-1}
\end{pmatrix} \notag\\
&\equiv \begin{pmatrix} a_n & b_n \\ c_n & d_n \end{pmatrix}.
\end{align}
Necessity: if $(a_{n-1}, b_{n-1})$ does not cover $\mathbb{R}^2\setminus\{0,0\}$, then $(c_n, d_n)$ cannot take on arbitrary values in $\mathbb{R}^2\setminus\{0,0\}$, thus $\mathbf{M}_n(\kappa_1,\dots,\kappa_{n})$ is not universal in $\text{Sp}(2, \mathbb{R})$.\\
Sufficiency: in the case of $c_n=a_{n-1}\neq 0$, $a_n$ can take on an arbitrary real value that is determined by $\kappa_n$.
Now $a_n$, $c_n=a_{n-1}\neq 0$, and $d_n=b_{n-1}$ take on arbitrary values, and $b_n$ is automatically determined from the condition $a_nd_n-b_nc_n=1$, as $c_n\neq 0$.
In the case when $c_n=a_{n-1}=0$, we have $d_n=b_{n-1}\neq 0$, and $\mathbf{M}_n(\kappa_1,\dots,\kappa_{n})$ has the form $\left(\begin{smallmatrix} 1/d_n & b_n \\ 0 & d_n \end{smallmatrix}\right)$;
$b_n=-d_{n-1}-\kappa_n b_{n-1}$ takes on an arbitrary value determined by $\kappa_n$, as $b_{n-1}\neq 0$. Q.E.D.

Using the above lemma, we can show that the minimum number of elementary steps that 
is required for universal one-mode Gaussian transformations is four.
Because there are three degrees of freedom (DOF) for $\text{Sp}(2, \mathbb{R})$, one might expect that three steps are sufficient. However, some measure-zero set of operations in $\text{Sp}(2, \mathbb{R})$ cannot be achieved with only three steps.
This is expressed by the following theorem.\\
{\it Theorem}:~In order to realize an arbitrary one-mode LUBO transformation through one-way computation over
CV, four elementary teleportation steps, involving quadratic phase gates and Fourier transforms, are necessary and sufficient.\\
{\it Proof}: The matrix representation for two steps is $M(\kappa_2)M(\kappa_1)=\left(\begin{smallmatrix} \kappa_2\kappa_1-1 & \kappa_2 \\ -\kappa_1 & -1 \end{smallmatrix}\right)\equiv\left(\begin{smallmatrix} a_2 & b_2 \\ c_2 & d_2 \end{smallmatrix}\right)$, thus when $b_2=0$, 
the parameter $a_2$ cannot take on a value other than $-1$.
As a consequence, $M(\kappa_3)M(\kappa_2)M(\kappa_1)=\left(\begin{smallmatrix} -\kappa_3\kappa_2\kappa_1+\kappa_3+\kappa_1 & -\kappa_3\kappa_2+1 \\ \kappa_2\kappa_1-1 & \kappa_2 \end{smallmatrix}\right)\equiv\left(\begin{smallmatrix} a_3 & b_3 \\ c_3 & d_3 \end{smallmatrix}\right)$ cannot have $d_3=0$ and $b_3\neq 1$;
hence three elementary steps $M(\kappa_3)M(\kappa_2)M(\kappa_1)$ are not universal for $\text{Sp}(2, \mathbb{R})$.

On the other hand,
$(a_3,b_3)=(-b_3\kappa_1+\kappa_3,-\kappa_3\kappa_2+1)$ does cover the whole range $\mathbb{R}^2\setminus\{0,0\}$, as follows.
The parameter $b_3$ takes on an arbitrary real value independent of $\kappa_1$.
In the case of $b_3\neq 0$, $a_3$ can then take on an arbitrary real value that is determined by $\kappa_1$.
In the case of $b_3=0$, $\kappa_3=1/\kappa_2$ takes on an arbitrary real value different from zero, and so does $a_3$.
As a result, using the above lemma,
four elementary steps $M(\kappa_4)M(\kappa_3)M(\kappa_2)M(\kappa_1)$ are (necessary and)
sufficient for universal one-mode Gaussian operations. Q.E.D.

We complete this discussion by presenting the explicit choice of parameters $\kappa_1,\dots,\kappa_4$.
The total matrix for four steps is,
\begin{align}
&M(\kappa_4)M(\kappa_3)M(\kappa_2)M(\kappa_1) \notag\\
&=\left(
\begin{smallmatrix}
\kappa_4\kappa_3\kappa_2\kappa_1-\kappa_4\kappa_3-\kappa_2\kappa_1-\kappa_4\kappa_1+1 &
\kappa_4\kappa_3\kappa_2-\kappa_4-\kappa_2 \\
-\kappa_3\kappa_2\kappa_1+\kappa_3+\kappa_1 &
-\kappa_3\kappa_2+1
\end{smallmatrix}\right). \label{EqDecomOneMode}
\end{align}
An arbitrary one-mode Gaussian operation represented by $M_{G(1)} = \left(\begin{smallmatrix} a & b \\c & d \end{smallmatrix}\right)\in\text{Sp}(2, \mathbb{R})$ can be decomposed into $M(\kappa_4)M(\kappa_3)M(\kappa_2)M(\kappa_1)$, as follows,
\begin{align}
\kappa_2=\dfrac{1-d}{\kappa_3},\, \kappa_3=c-d\kappa_1,\,\kappa_4=\dfrac{1-a+b\kappa_1}{\kappa_3},
\end{align}
where $\kappa_1$ is a free parameter which should be typically chosen such that $\kappa_3\neq 0$,
unless the numerators of $\kappa_2$ and $\kappa_4$ in the above equations are zero, for which $\kappa_3$ may become zero.
One simple example is the identity operation $M_{G(1)} = \left(\begin{smallmatrix} 1 & 0 \\ 0 & 1 \end{smallmatrix}\right)$, which corresponds to $\kappa_1=\dots =\kappa_4=0$.

As those operations which are not achievable through a three-step computation
are only a small subset of the whole set of Gaussian operations, 
one might just consider approximations infinitesimally close to them.
However, in the realistic case, this is not a good strategy,
because the squeezing of the ancilla cluster states will be finite.
In this case, the finite-squeezing-induced excess noise grows arbitrarily big
for three-step circuits that aim at sufficiently closely approximating otherwise
unachievable operations.
In the four-step case, however, such large excess noises are avoided, 
and furthermore, the extra degree of freedom can be exploited to minimize the excess noise.

\section{INPUT COUPLING THROUGH TELEPORTATION}\label{telepcoupling}
For Clifford one-mode one-way quantum computation, it is straightforward to
apply the results of the preceding section on universal
one-mode LUBO transformations directly to the most general scenario
where an arbitrary input quantum state is attached to the ancilla cluster state
through QND coupling. In this general case, the input state may have been already
processed and may correspond to the output of an earlier quantum computation.
A crucial question then is how to achieve this input coupling between a fragile
input quantum state and the ancilla cluster state in an efficient and practical way.
In this section, we address this issue.

Note that there is an essential difference between the QND couplings for the initial ancilla squeezed states and those that couple an input state to the ancilla state.
Arbitrary ancilla cluster (or graph) states can be built through linear optics
using beam splitters and offline single-mode squeezed states, as has been shown theoretically \cite{Peter07P}
and also demonstrated for some examples experimentally \cite{Yukawa08}.
Hence, as opposed to the actual input-cluster coupling,
the QND couplings for cluster generation can be effectively replaced by beam splitters.
Of course, there are situations when the input state may not be coupled to the cluster from the outside.
In principle, an arbitrary multi-mode state can be prepared as a subset of modes from a larger cluster state,
and, in this case, there is no need to prepare an independent input state before the cluster-state generation.
One may just prepare any desired state within the cluster and then proceed with the quantum computation.

Such a strategy, however, can be rather inefficient, especially, in a one-way computation
with Gaussian cluster states \cite{Detail}.
Furthermore, there might be situations in which the input coupling is necessary, for instance,
when an unknown state has been transmitted through a quantum channel and is to be further processed
through cluster computation.

Provided that efficient QND couplings are available, we may just prepare the ancilla cluster state offline,
and attach an input state to the cluster through QND coupling.
However, alternatively, we may also employ a nonlocal measurement for this input coupling.
A so-called Bell measurement, which is the two-mode measurement used in quantum teleportation
\cite{Braunstein98}, is the prime example for such a nonlocal measurement.
In the following, we will discuss this type of coupling for arbitrary input states through quantum teleportation.
In an optical realization, an important advantage is that the Bell measurement can be easily implemented with a beam splitter and two homodyne detections \cite{Furusawa98}.

\begin{figure}
\centering
\includegraphics[width=5cm,clip]{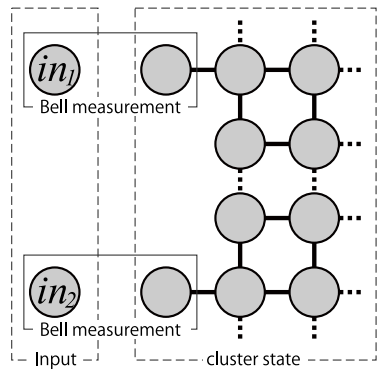}
\caption{Typical diagram for input coupling through quantum teleportation.
The input state is teleported into the cluster state by Bell measurements on
the input modes and the end nodes of the cluster state.
}
\label{FigAbstractBS}
\end{figure}

Figure~\ref{FigAbstractBS} shows a typical diagram of the input coupling,
in which a two-mode input state ia attached to the cluster through Bell measurements
on the input modes together with suitable `port' modes from the cluster state.
We will discuss only this situation, though there are many other possible configurations
that might complicate the problem.
Quantum teleportation with CV uses an Einstein-Podolsky-Rosen (EPR) type state as a resource.
The key point for our proposal is that a two-mode cluster state is an EPR state, up to a local Fourier transform.
Thus, the end nodes of a cluster state can be considered as an EPR pair, one half of which is connected to the rest of the cluster state through QND interactions.
By performing a Bell measurement on a single-mode input state and a cluster end node (the unconnected half of the
EPR state), the input state will be teleported to the connected side of the EPR pair,
located at the edge of the cluster state.
In the case that the input state is an $n$-mode (entangled) state, $n$ independent quantum teleportations using $n$ cluster end nodes would couple the input to the cluster, as depicted in Fig.~\ref{FigAbstractBS}.

We describe now the usual quantum teleportation protocol for teleporting an unknown input state into a two-mode ancilla cluster state (an EPR state).
The quantum correlations of the two-mode ancilla cluster state are $\hat{x}_2-\hat{p}_1\rightarrow 0$ and $\hat{x}_1-\hat{p}_2\rightarrow 0$.
We choose the linear beam splitter transformation for the Bell measurement as $\left(\begin{smallmatrix} \hat{a}_0^\prime \\ \hat{a}_1^\prime \end{smallmatrix}\right)=B_{0,1}\left(\begin{smallmatrix} \hat{a}_0 \\ \hat{a}_1 \end{smallmatrix}\right)=\tfrac{1}{\sqrt{2}}\left(\begin{smallmatrix} 1 & i \\ i & 1 \end{smallmatrix}\right)\left(\begin{smallmatrix} \hat{a}_0 \\ \hat{a}_1 \end{smallmatrix}\right)$,
where subscript `$0$' denotes the input mode, and the primes correspond the the output modes of the beam splitter.
The input-output relations for this beam splitter are
\begin{align}
\hat{x}_0^\prime&=(\hat{x}_0-\hat{p}_1)/\sqrt{2},& \hat{x}_1^\prime&=(\hat{x}_1-\hat{p}_0)/\sqrt{2}, \notag\\
\hat{p}_0^\prime&=(\hat{p}_0+\hat{x}_1)/\sqrt{2},& \hat{p}_1^\prime&=(\hat{p}_1+\hat{x}_0)/\sqrt{2}.
\end{align}
Measuring $\hat{x}_0^\prime$ and $\hat{x}_1^\prime$ is equivalent to a Bell measurement and leads to the standard quantum teleportation without any extra manipulation of the input state.

However, by modifying the nonlocal measurement basis compared to the Bell basis,
this teleportation does not only couple an input state to the cluster, but it also manipulates the input state
correspondingly.
With the above beam splitter coupling and subsequent homodyne measurements,
the possible operations are Gaussian, as we will see below.
The phases of the homodyne detections are expressed by $\theta_0$ and $\theta_1$, i.e.,
the observables $\hat{x}_0^\prime\cos\theta_0+\hat{p}_0^\prime\sin\theta_0$ and $\hat{x}_1^\prime\cos\theta_1+\hat{p}_1^\prime\sin\theta_1$ will be measured.
The resulting teleportation is associated with the following transformation:
\begin{align}
\begin{pmatrix}
\hat{x}^\prime \\
\hat{p}^\prime
\end{pmatrix}
&=
\begin{pmatrix}
\frac{\cos\theta_+}{\cos\theta_-} & \frac{\sin\theta_-+\sin\theta_+}{\cos\theta_-} \\
\frac{\sin\theta_--\sin\theta_+}{\cos\theta_-} & \frac{\cos\theta_+}{\cos\theta_-}
\end{pmatrix}
\begin{pmatrix}
\hat{x} \\
\hat{p}
\end{pmatrix} \notag\\
&\equiv M_\text{tel}(\theta_+, \theta_-)
\begin{pmatrix}
\hat{x} \\
\hat{p}
\end{pmatrix},
\end{align}
where $\theta_\pm=\theta_0\pm\theta_1$.
The standard teleportation (identity transfer) corresponds to the case $\theta_0=\theta_1=0$.
In the case $\theta_-=\pi/2+n\pi$, $n\in\mathbb{Z}$, the teleportation is not successful, because one quadrature of the input state is perfectly measured and the information of the orthogonal quadrature is lost; correspondingly, the elements of the matrix $M_\text{tel}(\theta_+, \theta_-)$ go to infinity.
In the following, we assume $\cos\theta_->0$.
For the case of $\cos\theta_-<0$, we can redefine $\theta_+^\prime=\theta_+ +\pi$ and $\theta_-^\prime=\theta_- +\pi$, which results in identical transformations, i.e., $M_\text{tel}(\theta_+, \theta_-)=M_\text{tel}(\theta_+^\prime, \theta_-^\prime)$, and $\cos\theta_-^\prime>0$.

This seemingly complicated transformation can be intuitively understood by considering the following two cases separately.
On one hand, in the case that the two local measurement bases $\theta_0$ and $\theta_1$ are rotated in the same direction and by the same amount, i.e., $\theta_+\neq 0$ and $\theta_-=0$, we obtain a phase space rotation,
\begin{equation}
M_\text{tel}(\theta_+, 0)=\begin{pmatrix} \cos\theta_+ & \sin\theta_+ \\ -\sin\theta_+ & \cos\theta_+ \end{pmatrix}=R(-\theta_+).
\end{equation}
On the other hand, in the case that the two local measurement bases $\theta_0$ and $\theta_1$ are rotated in opposite directions by the same amount, i.e., $\theta_+=0$ and $\theta_-\neq 0$, squeezing will occur along the 45$^\circ$ direction,
\begin{align}
&M_\text{tel}(0, \theta_-) \notag\\
&=\begin{pmatrix} \frac{1}{\cos\theta_-} & \frac{\sin\theta_-}{\cos\theta_-} \\ \frac{\sin\theta_-}{\cos\theta_-} & \frac{1}{\cos\theta_-} \end{pmatrix} \notag \\
&=\frac{1}{2} \begin{pmatrix} 1 & -1 \\ 1 & 1\end{pmatrix}
\begin{pmatrix} \frac{1+\sin\theta_-}{\cos\theta_-} & 0 \\ 0 & \frac{1-\sin\theta_-}{\cos\theta_-} \end{pmatrix}
\begin{pmatrix} 1 & 1 \\ -1 & 1\end{pmatrix} \notag\\
&=R(\pi/4)S(r(\theta_-))R(-\pi/4) \notag\\
&=\begin{pmatrix}\cosh r(\theta_-) & \sinh r(\theta_-) \\ \sinh r(\theta_-) & \cosh r(\theta_-) \end{pmatrix},
\end{align}
where $S(r)=\left(\begin{smallmatrix} \exp(r) & 0 \\ 0 & \exp(-r) \end{smallmatrix}\right)$ describes a squeezing operation, with $r>0$ corresponding to $p$-squeezing and $r<0$ corresponding to $x$-squeezing.
The squeezing parameter $r(\theta_-)$ is determined by $\tanh r(\theta_-) = \sin\theta_-$.
In the case of general $\theta_+$ and $\theta_-$, the resulting operation is a combination of the above two cases:
\begin{align}
&M_\text{tel}(\theta_+, \theta_-) \notag\\
&=M_\text{tel}(\theta_+/2, 0)M_\text{tel}(0, \theta_-)M_\text{tel}(\theta_+/2, 0) \notag\\
&=R(-\theta_+/2+\pi/4)S(r(\theta_-))R(-\theta_+/2-\pi/4).
\end{align}
This is a 45$^\circ$-tilted squeezing operation sandwiched by rotations at an angle of $\theta_+/2$.
In the next section, we will use this result to describe a general one-mode LUBO transformation
with teleportation-based input coupling.

\section{ONE-MODE LUBO WITH TELEPORTATION-BASED COUPLING}\label{onemodeLUBOwtelep}
In the case that the relative phase at the beam splitter (for teleportation) may be changed arbitrarily, the teleportation protocol alone is sufficient to realize arbitrary one-mode Gaussian operations.
We shall briefly explain this approach which partly violates the rules of one-way cluster
protocols, as the state manipulation depends on the choice of nonlocal measurement bases (projections
onto which require corresponding adjustments of the beam splitter coupling for teleportation).

It is known that an arbitrary matrix in $\text{Sp}(2, \mathbb{R})$ can be decomposed as \cite{Braunstein05}:
\begin{equation}
M_{G(1)}=R(\phi_1)S(\xi)R(\phi_2).
\end{equation}
The corresponding LUBO transformation of the annihilation operator $\hat{a}$ is $\hat{a}^\prime=\mu\hat{a}+\nu\hatd{a}$ where $\mu=\exp [-i(\phi_1+\phi_2)]\cosh\xi$ and $\nu=\exp [-i(\phi_1-\phi_2)]\sinh\xi$.
Now the 2$\times$2 matrix representation of the generalized teleportation with an extra phase rotation
beforehand is $M_\text{tel}(\theta_+,\theta_-)R(\theta_{in})=R(-\theta_+/2+\pi/4)S(r(\theta_-))R(-\theta_+/2-\pi/4+\theta_{in})$.
As a result, an arbitrary one-mode Gaussian operation can be achieved with the appropriate choice of $\theta_+$, $\theta_-$, and $\theta_{in}$.

\begin{figure}[b]
\centering
\includegraphics[width=7.5cm,clip]{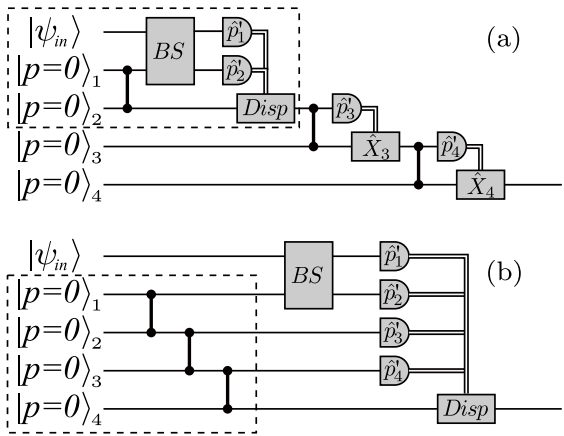}
\caption{(a)~Input coupling scheme through quantum teleportation (dashed box)
followed by two elementary gates, allowing for arbitrary one-mode Gaussian operations.
(b)~An equivalent circuit to (a). The circuit enclosed by the dashed box in (b) shows a four-mode
linear cluster state. Measurements and feedforwards can all be put at the end of the circuit.}
\label{FigFourStepBS}
\end{figure}

For the more interesting case when we stick to the rules of cluster computation
(i.e., we consider only the DOF of the local measurement bases),
the relative phase at the beam splitter must be fixed, and so an additional two-step quadratic phase gate followed by Fourier transforms is needed for an arbitrary one-mode Gaussian operation.
In other words, when we replace the QND coupling between the input state and the cluster state (Fig.~\ref{FigFourStepTele}) by a beam splitter interaction (Fig.~\ref{FigFourStepBS}), the required number of modes of the linear ancilla cluster state remains four.

In order to show this, we use again the lemma proven above.
We substitute $\theta_-$ by $r= \arctanh \sin\theta_-$, omit the subscript $+$ in $\theta_+$, and rewrite $M_\text{tel}(\theta_+, \theta_-)$ as $M_\text{tel}^\prime(\theta, r)$,
\begin{align}
&M_\text{tel}^\prime(\theta, r) \notag\\
&=
\begin{pmatrix}
\cos\theta\cosh r & \sin\theta\cosh r +\sinh r \\
-\sin\theta\cosh r + \sinh r & \cos\theta\cosh r
\end{pmatrix} \notag\\
&\equiv
\begin{pmatrix}
a_T & b_T \\ c_T & d_T \label{EqaTbT}
\end{pmatrix}.
\end{align}

\begin{figure}
\centering
\includegraphics[width=6cm,clip]{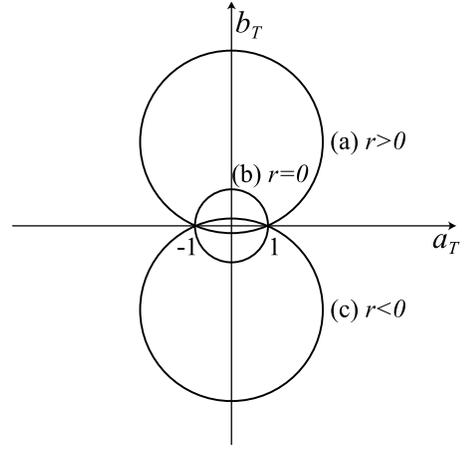}
\caption{The loci of $(a_T,b_T)$ in Eq.\eqref{EqaTbT}.
(a), (b), and (c) show the cases of $r>0$, $r=0$, and $r<0$, respectively.
}
\label{FigCircles}
\end{figure}

Let us consider the loci of $(a_T,b_T)$ in the $\mathbb{R}^2$ plane.
When the squeezing parameter $r$ is fixed, the locus of $(a_T,b_T)=(\cos\theta\cosh r,\sin\theta\cosh r +\sinh r)$ is a circle, the center of which is $(0,\sinh r)$, intersecting the $a_T$-axis in points $(\pm 1,0)$ regardless of $r$ (Fig.~\ref{FigCircles}).
Thus, the set of unreachable points of $(a_T,b_T)$ in $\mathbb{R}^2\setminus(0,0)$ is $N=\{(a,0)| a\neq\pm 1,0\}$.
As $(a_T,b_T)$ in $M_\text{tel}^\prime(\theta, r)$
does not cover the whole range of $\mathbb{R}^2\setminus(0,0)$, using the lemma, we conclude that
an additional elementary step,
$M(\kappa_3)$, following $M_\text{tel}^\prime(\theta, r)$ is not enough
for arbitrary Gaussian one-mode operations.
However, teleportation-based coupling followed by an additional elementary step,
$M(\kappa_3)M_\text{tel}^\prime(\theta, r)=\left(\begin{smallmatrix} -c_T-\kappa_3 a_T & -d_T-\kappa_3 b_T \\ a_T & b_T \end{smallmatrix}\right)\equiv\left(\begin{smallmatrix} a_3 & b_3 \\ c_3 & d_3 \end{smallmatrix}\right)$ does 
allow for arbitrary real values of $(a_3, b_3)$ except $(0,0)$; thus, using the lemma,
yet another additional step $M(\kappa_4)$ added to $M(\kappa_3)M_\text{tel}^\prime(\theta, r)$
does the trick and achieves Gaussian one-mode universality.

In order to prove the above statement, we have to show that $(a_3, b_3)$ covers $\mathbb{R}^2\setminus(0,0)$.
For $\kappa_3=0$, the unreachable points of $(a_3,b_3)$ in $\mathbb{R}^2\setminus(0,0)$ are those of $(-c_T,-d_T)$;
the corresponding set is $N^\prime=\{(0,b)| b\neq\pm 1,0\}$, using the same arguments as for $(a_T,b_T)$.
Therefore, by showing that an arbitrary point $(0,b)\in N^\prime \subset \mathbb{R}^2\setminus(0,0)$ is attainable
for some nonzero $\kappa_3$, the proof is complete.
We show this as follows:
for $a_3=0$, $\cos\theta$ should be nonzero, and $\kappa_3=-c_T/a_T=(\sin\theta\cosh r-\sinh r)/\cos\theta\cosh r$. 
Then $b_3$ is calculated as $b_3=-1/\cos\theta\cosh r$, which takes on an arbitrary real value other than zero. Q.E.D.

Below we give the explicit choice of the measurement bases for the implementation of a particular Gaussian operation expressed as $M_{G(1)}=\left(\begin{smallmatrix} a & b \\ c & d \end{smallmatrix}\right)$ through teleportation-based coupling followed by 
two additional elementary steps.
The two parameters $\theta_+$ and $\theta_-$ (the measurement bases of the teleportation coupling)
are determined only from the matrix elements $c$ and $d$, so that $c\sin\theta_+-d\cos\theta_+=\cos\theta_--c\sin\theta_-$.
Then the other parameters are given by $\kappa_3=-(d\cos\theta_-+\cos\theta_+)/(\sin\theta_++\sin\theta_-)$ and $\kappa_4=-[a+(\cos\theta_+/\cos\theta_-)]/c$.
A solution of these equations is,
\begin{align}
\cot\theta_1&=\frac{1-d}{2c-(1+d)\cot\theta_0}, \notag\\
\kappa_3&=c-(1+d)\cot\theta_0, \notag\\
\kappa_4&=\frac{1-a+b\cot\theta_0}{c-d\cot\theta_0},
\end{align}
where $\theta_0$ is a free parameter, which can be utilized to minimize excess noises, as described above.
Note that the problem of zero denominators in the intermediate expressions of $\kappa_3$ and $\kappa_4$ is 
avoided in the final forms for a suitable choice of $\theta_0$.

\section{UNIVERSAL MULTI-MODE LUBO}\label{multimodeLUBO}
In the remainder of this paper, as a final issue, we discuss arbitrary multi-mode Gaussian operations
(general multi-mode LUBO transformations).
We will present an explicit way to implement any multi-mode Gaussian operation using
a finite-sized cluster state and homodyne measurements on it.

The one-way two-mode entangling gate proposed previously \cite{Menicucci06} corresponds to
a QND interaction with unit gain (the same gate that is used to create the ancillary,
unweighted cluster/graph state).
In order to transfer this gate onto a two-mode input state,
the state has to propagate through a two-dimensional cluster state.
Even though, in principle, sufficient for achieving universality with CV (when supplemented by
arbitrary single-mode gates), the use of a single fixed-gain two-mode interaction gate
for multi-mode transformations is rather awkward, as arbitrary two-mode beam splitter interactions
have proven very powerful for multi-mode linear optics \cite{Reck94}.

\begin{figure}
\centering
\includegraphics[width=7.5cm,clip]{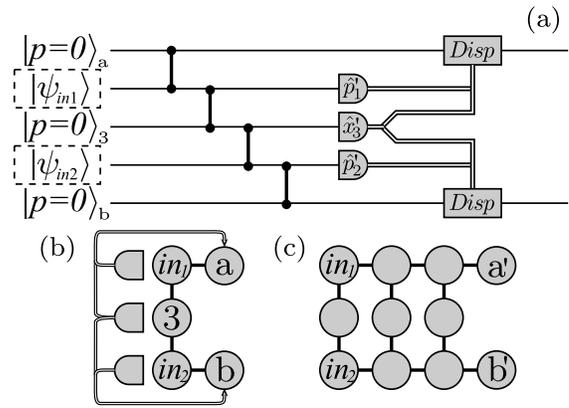}
\caption{(a)~Three-mode connection gate. Mode $in1$ and mode $in2$ are the input modes,
while modes $a$, $b$, and 3 are ancilla modes.
The input state may be an entangled state.
Measurement results on mode 1 and mode 3 are added (electronically) and used to displace mode $a$,
while measurement results on mode 2 and mode 3 are added (electronically) and used to displace mode $b$.
(b)~A graph representation of (a).
(c) shows a three-step three-mode connection gate in graph representation.
A phase-free beam splitter can be implemented using this configuration.
Measurements and feedforwards are omitted in (c).
}
\label{FigThreeModeConnect}
\end{figure}

Here, instead of a fixed-gain interaction, we propose another type of interaction,
referred to as a three-mode connection gate.
Its configuration is shown in Fig.~\ref{FigThreeModeConnect}(a),(b).
In this scheme, one ancilla mode would function as a kind of controller of the interaction gain.

In Fig.~\ref{FigThreeModeConnect}, mode $in1$ and mode $in2$ represent the input modes (in an arbitrary,
potentially entangled two-mode state),
while mode 3, mode $a$, and mode $b$ are ancilla squeezed vacuum modes.
Mode 3 plays the role of a controller of the interaction;
mode $a$ and mode $b$ are the end points for the propagation of quantum information from mode $in1$ and mode $in2$, respectively.
As before, links between cluster nodes represent QND couplings.

The measured variable at mode 3 is $\hat{x}_3^\prime=\hatd{O}(\hat{p}_3)\hat{x}_3\hat{O}(\hat{p}_3)=\hat{x}+\eta_3\hat{p}$ where $\hat{O}(\hat{p}_3)=e^{-i\eta_3\hat{p}_3^2}$.
The resulting interaction is $\exp[i\eta_3(\hat{x}_1+\hat{x}_2)^2]$.
On the other hand, the measurements on mode $in1$ and mode $in2$
correspond to the quadratic phase gates $\exp(i\kappa_1\hat{x}_1^2)$ and $\exp(i\kappa_2\hat{x}_2^2)$, respectively,
followed by Fourier transforms.
Note that the above three operators, $\exp[i\eta_3(\hat{x}_1+\hat{x}_2)^2]$, $\exp(i\kappa_{1}\hat{x}_1^2)$,
and $\exp(i\kappa_2\hat{x}_2^2)$, all commute.

As a result, by combining these three measurements, an arbitrary two-mode operation is achieved
whose Lie algebra is quadratic with regard to the position operator $\hat{x}$, i.e., $\hat{O}(\hat{x}_1,\hat{x}_2)=\exp(c_1\hat{x}_1^2+c_I\hat{x}_1\hat{x}_2+c_2\hat{x}_2^2)$.
The subsequent Fourier transform effectively swaps the roles of $\hat{x}$ and $\hat{p}$.
Thus, by cascading such three-mode connection gates, as illustrated in Fig.~\ref{FigThreeModeConnect}(c),
the two modes effectively interact subsequently with alternating quadratures
$\hat{x}$ and $\hat{p}$ for every single step.
Hence an $m$-time cascaded interaction may be written as,
\begin{align}
\hat{F}_2\dots\hat{F}_2 \hat{O}_m(\cdot)\dots\hat{O}_2(\vect{\hat{p}})\hat{O}_1(\vect{\hat{x}}),
\end{align}
where $F_2$ is a two-mode Fourier transform, and $\vect{\hat{x}} = (\hat{x}_1, \hat{x}_2)$, $\vect{\hat{p}} = (\hat{p}_1, \hat{p}_2)$.

Note that any interaction can be suppressed by setting $\eta_3=0$ such that interactions may be only applied
whenever they are needed for a fixed cluster state.
The 4$\times$4 matrix representation of the connection gate is,
\begin{align}
\begin{pmatrix}
\hat{x}_1^\prime \\
\hat{x}_2^\prime \\
\hat{p}_1^\prime \\
\hat{p}_2^\prime
\end{pmatrix}
\!=\!
\begin{pmatrix}
0_2 & -1_2 \\
1_2 & 0_2
\end{pmatrix}\!\!
\begin{pmatrix}
1 & 0 & 0 & 0 \\
0 & 1 & 0 & 0 \\
\kappa_1\!-\!\eta_3 & -\eta_3 & 1 & 0 \\
-\eta_3 & \kappa_2\!-\!\eta_3 & 0 & 1
\end{pmatrix}\!\!
\begin{pmatrix}
\hat{x}_1 \\
\hat{x}_2 \\
\hat{p}_1 \\
\hat{p}_2
\end{pmatrix}
, \label{EqThreeModeConnectGate}
\end{align}
where $0_2$ is a $2\times 2$ zero matrix and $1_2$ is a $2\times 2$ identity matrix;
$F_2=\left(\begin{smallmatrix} 0_2 & -1_2 \\ 1_2 & 0_2 \end{smallmatrix}\right)$ is the matrix representation of the two-mode Fourier transform.

To complete the discussion on arbitrary Gaussian multi-mode transformations,
we shall use the well-known decomposition of multi-mode Gaussian operations, usually referred to as Bloch-Messiah reduction \cite{Braunstein05}.
An arbitrary $n$-mode Gaussian operation $\hat{G}$, whose DOF are $2n^2+n$, is decomposed into the form $\hat{U}\hat{S}\hat{V}$, where $\hat{U}$ and $\hat{V}$ correspond to passive linear-optics circuits with $n^2$ DOF coming from beam splitters (with some fixed phase) and single-mode phase shifters; $\hat{S}$ represents single-mode squeezers applied to each mode.

The phase shifters and squeezers are one-mode operations which are realizable using at most four ancilla modes, as discussed in detail before.
Thus, provided an explicit implementation of a phase-free beam splitter with arbitrary reflectivity $R$ is given, we can conclude that any multi-mode Gaussian operation is achievable with our specifically shaped, finite-sized cluster (where our implementation may be suboptimal).
A decomposition of the linear-optics circuits $\hat{U}$ and $\hat{V}$ into beam splitters and phase shifters requires at most $n(n-1)/2$ phase-free beam splitters and $n(n+1)/2$ phase shifters \cite{Reck94}.
Thus, the number of ancilla modes required for this implementation is quadratic in the number of input modes $n$.
It is now worth noting that the number of DOF of $\text{Sp}(2n, \mathbb{R})$ is $2n^2+n$, corresponding to a minimum size of a cluster state
for universal multi-mode Gaussian operations also quadratic with regard to $n$.
Hence our one-way scheme with a total cluster state of size $\sim n^2$
(using a supply of four-mode linear subclusters and the corresponding subclusters for three-mode connection gates)
would provide an efficient realization of universal multi-mode LUBO transformations.

Finally, in order to establish the link between the three-mode connection gates
and phase-free beam splitters,
let us define a phase-free beam splitter with intensity reflectivity $R$,
\begin{align}
\!\!\!
\begin{pmatrix}
\hat{x}_1^\prime \\
\hat{x}_2^\prime \\
\hat{p}_1^\prime \\
\hat{p}_2^\prime
\end{pmatrix}
=
\begin{pmatrix}
M_R & 0_2 \\
0_2 & M_R
\end{pmatrix}
\begin{pmatrix}
\hat{x}_1 \\
\hat{x}_2 \\
\hat{p}_1 \\
\hat{p}_2
\end{pmatrix}, \notag\\
M_R=
\begin{pmatrix}
\sqrt{R} & \sqrt{1-R} \\
\sqrt{1-R} & -\sqrt{R}
\end{pmatrix}.
\end{align}
Note that $M_R^2=1_2$.
We have the following relation,
\begin{align}
\begin{pmatrix}
M_R & 0_2 \\
0_2 & M_R
\end{pmatrix}
=
\left[
\begin{pmatrix}
0_2 & -1_2 \\
1_2 & 0_2
\end{pmatrix}
\begin{pmatrix}
1_2 & 0_2 \\
M_R & 1_2
\end{pmatrix}
\right]^3
\equiv
M_I^3.
\end{align}
The transformation $M_I$ is achieved using a three-mode connection gate, choosing the three parameters $\kappa_1, \kappa_2$, and $\eta_3$ in the following way,
\begin{align}
\kappa_1&=\sqrt{R}-\sqrt{1-R}, \notag\\
 \kappa_2&=-\sqrt{R}-\sqrt{1-R}, \notag\\
\eta_3&=-\sqrt{1-R}.
\end{align}
Therefore, a phase-free beam splitter with an arbitrary reflectivity $0\le R\le 1$ can be implemented through
a three-step three-mode connection gate. This would require in total nine ancilla modes.

\section{CONCLUSION}\label{conclusion}

In conclusion, we have described an explicit implementation for arbitrary one-mode and
multi-mode linear unitary Bogoliubov (LUBO) transformations (Gaussian operations)
in the framework of one-way computation over continuous variables
using Gaussian cluster states and homodyne measurements.
We have shown that an ancillary, linear four-mode cluster state is a necessary and
sufficient resource for universal one-mode Gaussian operations.
We have also presented a strategy for multi-mode Gaussian operations,
where beam splitter interactions are used as the sole multi-mode operation.
Arbitrary (phase-free) beam splitters can be realized in a measurement-based
one-way scheme through so-called three-mode connection gates consuming
one ancilla three-mode cluster per gate. Every beam splitter requires
three such three-mode connection gates, so nine ancilla modes in total.

Most importantly, our scheme scales quadratic with the number of input modes
such that an ancilla cluster state of size at most quadratic in the number of
input modes is sufficient. This scaling coincides with the scaling of the number
of elementary optical gates (phase shifters, beam splitters, and squeezers)
needed for a circuit implementation of general LUBO transformations.
We leave a possible optimization of our multi-mode cluster-based
scheme for future research.

Towards actual experimental demonstrations of the results derived here,
we discussed some simplifications for coupling arbitrary input states
to a given cluster state. Our simplified scheme would be based on
standard quantum teleportation instead of the more expensive QND coupling.
Remarkably, eventually, the coupling QND gate may just be replaced by
a fixed beam splitter, as already through our generalized teleportation scheme,
it is possible to manipulate and process the input state to some extent.

One big strength of our scheme is as follows. As it is well-known how
to generate arbitrary cluster/graph states using linear optics,
by employing the present scheme, one may now
perform a {\it general} multi-mode LUBO transformation on an {\it arbitrary}
multi-mode input state (including fragile non-Gaussian states)
in an efficient, solely measurement-based fashion.
All potentially inefficient, optical interactions (such as online squeezing)
would be done beforehand offline for the resource cluster state.
Although efficient multi-mode LUBO transformations are now, in principle,
accessible even for non-Gaussian input states, in a realistic scheme,
only an approximate, finitely squeezed ancilla cluster state could be used.
Therefore, the resulting LUBO transformations would become imperfect,
depending on the initial squeezing level. Apart from utilizing
new experimental schemes with further increasing squeezing levels,
one possibility to address the finite-squeezing issue may be in form
of some kind of error correction such as postselection \cite{Menicucci06}
or redundant encoding \cite{Peter07P}.

\section{ACKNOWLEDGMENTS}
This work was partly supported by SCF, GIA, G-COE, and PFN commissioned by the MEXT of Japan,
the Research Foundation of Opt-Science and Technology, and SCOPE program of the MIC of Japan.
P.v.L. acknowledges support from the Emmy Noether programme of the DFG in Germany.

\thispagestyle{empty}

\renewcommand{\refname}{\vspace{-1cm}}


\begin{thebibliography}{00}
\bibitem{detail2} We use `mode' and `qumode' interchangeably throughout.
\bibitem{Detail}
For example, consider the case when a coherent state is required as an input state.
Since ancilla cluster states are made of squeezed vacuum states,
in order to prepare a coherent state within the cluster, a reverse squeezing operation is required.
The required squeezing level of this operation would become
infinite in the ideal limit with cluster states built from infinitely squeezed states.
\end{thebibliography}
\end{document}